\documentclass[twocolumn]{article}

\usepackage{graphicx}
\usepackage{amsmath}
\usepackage{txfonts}

\usepackage{setspace}  
\usepackage{abstract}  

\usepackage{color} 
\usepackage{ulem}  
\usepackage{multirow} 

\newcommand{\micron}{$\mu$m}
\newcommand{\pcm}{cm$^{-1}$}
\newcommand*\arcmin{\ensuremath{^\prime}}
\newcommand*\arcsec{\ensuremath{^{\prime\prime}}}
\def\tablefoot#1{$^\mathrm{#1}$}
\definecolor{orange}{rgb}{0.8,0.4,0.0}
\begin{document}


\title{Astronomical imaging Fourier spectroscopy at far-infrared wavelengths}

\author{{\bf David A. Naylor, Brad G. Gom, Matthijs H.D. van der Wiel, and Gibion Makiwa} \\
\small Institute for Space Imaging Science, Department of Physics and Astronomy, University of Lethbridge, Lethbridge, AB T1K 3M4, Canada
} 

\date{\small To be published in {\bf Canadian Journal of Physics}. \\ Received: 2012 December 21; revision received: 2013 February 6; accepted: 2013 February 12.}

\twocolumn[
\maketitle

\begin{onecolabstract}
\noindent
The principles and practice of astronomical imaging Fourier transform spectroscopy (FTS) at far-infrared wavelengths are described. The Mach-Zehnder interferometer design has been widely adopted for current and future imaging FTS instruments; we compare this design with two other common interferometer formats. Examples of three instruments based on the Mach-Zehnder design are presented. The techniques for retrieving astrophysical parameters from the measured spectra are discussed using calibration data obtained with the {\it Herschel}-SPIRE instrument. The paper concludes with an example of imaging spectroscopy obtained with the SPIRE FTS instrument.  \\
{\bf PACS}: {95.55.-n (Astronomical and space-research instrumentation)}
\vspace{2em}
\end{onecolabstract}

]  

\section{Introduction	}

It is now well established that approximately one half of the radiant energy emitted by the universe falls in the far-infrared and submillimetre spectral range (30--1000~\micron) \cite{hauser2001} due to two principal reasons. The first is that sources in the distant universe, ultra luminous infrared galaxies in the local universe (ULIRG), or protostars in our own galaxy are often shrouded in dust and gas. The dust efficiently scatters and absorbs shorter wavelength radiation from the stellar photospheres. The absorbed energy is subsequently re-radiated at longer wavelengths, both as continuum emission from the dust and as line emission from the ions, atoms and molecules present in the emitting region. In the case of ULIRGs, up to 99\% of the radiant energy can occur in the far-infrared \cite{clements2010}. The second reason is that distant galaxies counter intuitively do not decrease in brightness with increasing redshift as $(1+z)^{-2}$ because the emission from the galaxy shifts into the infrared spectral range, an effect known as the negative K correction \cite{blain1999}, which compensates for cosmological dimming for redshifts out to $z$ $\sim$ 10 at far-infrared wavelengths.

A detailed understanding of the processes underlying star formation remains as one of the outstanding questions in modern astrophysics. Since star formation is intimately linked with planetary formation, this has fundamental significance not only for understanding our immediate galactic environment, but also in the study of the early universe. Thus, while observations at optical wavelengths provide important information on the formation of stars and galaxies, a detailed understanding of the physical processes at play requires complementary observations in the far-infrared. Unfortunately, most of this spectral region is inaccessible from the ground due to absorption by the Earth's atmosphere, which necessitates the use of space borne instrumentation. Moreover, since the far-infrared photon energy is $<10^{-2}$ eV, the instruments must be cooled, typically to liquid helium temperatures, in order to minimize the effects of thermal emission from the instruments themselves, which would otherwise dominate the weak astronomical signal. 

In the first infrared space telescopes (Infrared Astronomical Satellite (IRAS) \cite{neugebauer1984}, Infrared Space Observatory (ISO) \cite{kessler1996} and AKARI \cite{murakami2007}), cooling was achieved by placing the entire telescope in a liquid helium filled cryostat, limiting the diameters of primary mirrors to 60 cm. While these pioneering missions provided our first view of the far-infrared universe, their small apertures resulted in relatively low spatial resolution. The {\it Herschel} Space Observatory \cite{pilbratt2010}, launched by the European Space Agency in 2009, broke this barrier by employing a 3.5 m diameter passively cooled primary mirror located outside of the instrument payload, the instrument suite being cooled to $\sim$4~K by an on board supply of liquid helium. This design provides a major advance in spatial resolution and sensitivity, however, the sensitivity remains limited by the photon noise from the relatively warm ($\sim$80~K), albeit low emissivity telescope. Gains in sensitivity of two orders of magnitude will be realized by actively cooling larger aperture telescopes as proposed for the Japanese Space Agency led SPICA mission \cite{swinyard2009}.

\section{Fourier Spectroscopy}
While photometric observations in the far-infrared provide morphological information on the structure of astronomical sources, the composition and physical conditions of the sources can only be determined from spectral imaging observations. While different spectroscopic techniques have been explored to this end, imaging Fourier transform spectrometers (iFTS), have been widely adopted in the far-infrared. With their high throughput, broad spectral coverage and variable resolution, coupled with their well-defined instrumental line shape and intrinsic wavelength and intensity calibration, Fourier spectrometers have a long history in astronomical research \cite{connes1970}. Furthermore, imaging spectroscopy is relatively easily achieved with Fourier transform spectrometers by the addition of an array detector, providing a spectrum at each spatial coordinate viewed by a pixel in the array. 

\begin{figure}
	\resizebox{\hsize}{!}{\includegraphics{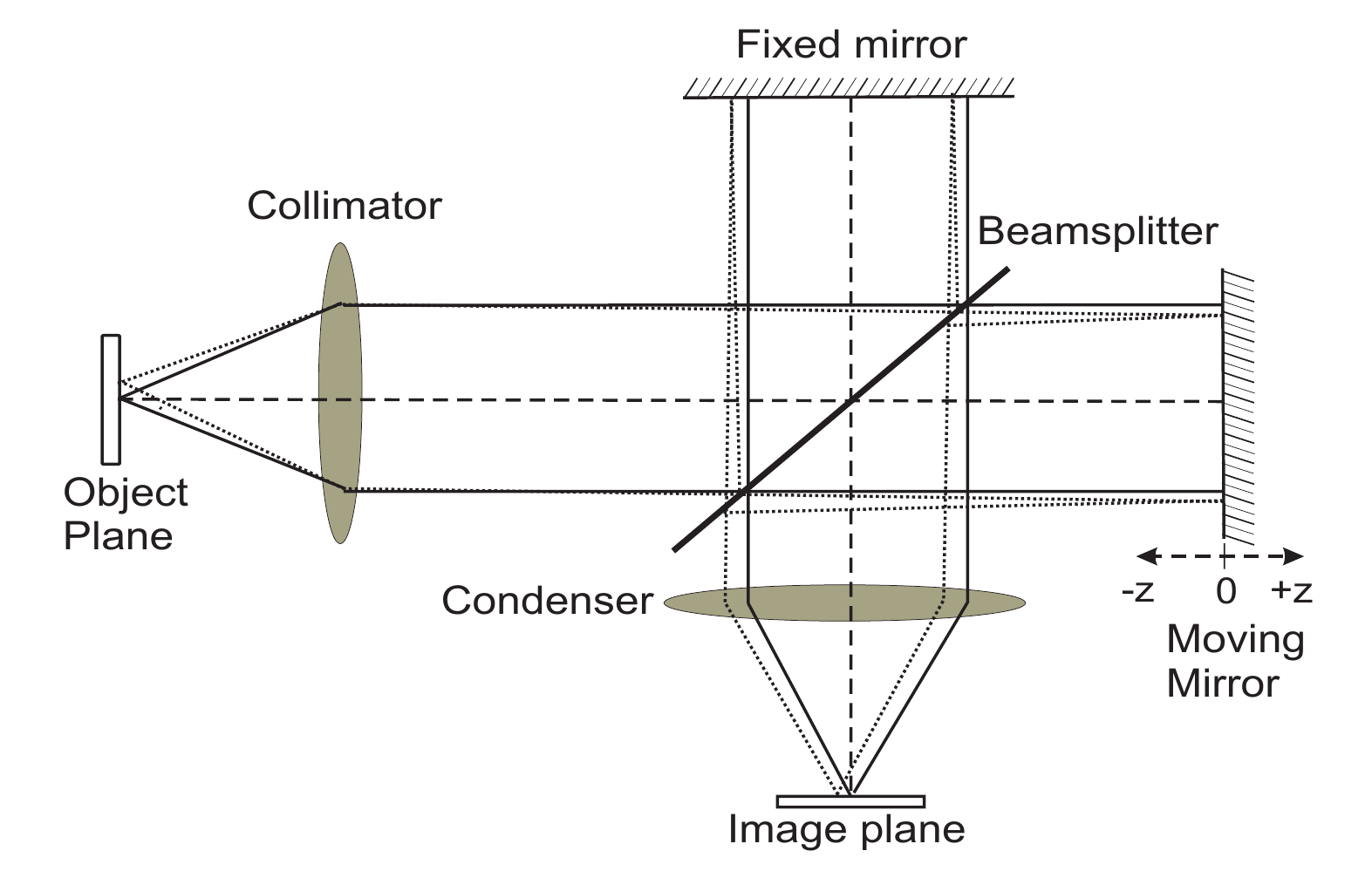}}
	\caption{Optical configuration of the Michelson interferometer. The astronomical signal from the object plane (left) is propagated through the interferometer as described in the text, with the optical path difference $z$ created by the moving mirror (right). The detector array is located in the image plane (bottom).}
	\label{fig:michelson}
\end{figure}

In the simplest type of Fourier spectrometer, the Michelson interferometer, shown in Figure~\ref{fig:michelson}, the incoming beam of light (e.g. from the telescope) is divided into two beams of equal intensity by a beamsplitter. After reflection from a fixed and a moving mirror, the beams recombine at the beamsplitter and are brought to a focus on the detector.

The signal recorded by the detector as a function of the optical path difference, $z$, between the recombining beams is known as the interferogram. In the case of a bolometer, the signal is proportional to the square of the electric field. The interferogram represents the autocorrelation of the electric field, $E(\sigma)$, of the incident radiation and, in the ideal case, is given by the cosine Fourier transform of the  spectrum, $B(\sigma)$, where $\sigma$ is the frequency expressed in wavenumbers (\pcm)
\begin{equation}
I(z) = \int\limits_{-\infty}^\infty B(\sigma) \cos(2 \pi z \sigma) \mathrm{d}\sigma
\end{equation}
where $B(\sigma)$ is proportional to $|E(\sigma)|^2$. It can be seen that in the ideal case the interferogram is symmetrical about the zero optical path difference position ($z=0$). In principle the spectrum is recovered by computing the cosine transformation of the interferogram. In practice, phase errors that arise from optical or electronic effects \cite{davis2001} produce interferograms that are no longer symmetrical about $z=0$, and therefore require the use of the complex Fourier transform.
\begin{equation}
B(\sigma) = \int\limits_{-\infty}^\infty I(z) \mathrm{e}^{-i 2 \pi \sigma z} \mathrm{d}z
\end{equation}
Thus, while in principle the design of an FTS is relatively straightforward, involving only two interfering beams, obtaining the spectrum requires sophisticated mathematical analysis. 

Regardless of their design, all Fourier transform spectrometers possess two input ports and two output ports. In the case of a Michelson interferometer \cite{hecht2008}, the input and output ports are superimposed. Two consequences result: the first is that 50\% of the incident radiation is returned to the source, the second is that radiation emitted by the detector assembly itself can be modulated by the interferometer and return to the detector where it is subsequently detected. While this effect is of little or no concern in the optical and near-infrared, since typical instrument operating temperatures produce no emission at these wavelengths, this is not the case in the far-infrared. For example, an instrument cooled to 30K has its peak emission at 100~\micron. For this reason, Fourier transform spectrometer designs that separate the two input and two output ports are required for operation at long wavelengths. 

Historically, the Martin-Puplett (MP) polarizing interferometer \cite{martin1970} has been the spectrometer of choice for astronomical spectroscopy at far-infrared wavelengths. There are two principal reasons: first the modulation efficiency of a polarizing beamsplitter is both high and uniform over a wide spectral range. Secondly, as discussed above, the polarizing interferometer separates the two input and two output ports. A calibration source may be placed at one input port to provide absolute intensity calibration, or the two input ports can be configured to allow differential measurements, for example when one is trying to detect a weak astronomical signal in the presence of a large atmospheric emission component. Since the interferometric signals at the two output ports are complementary, they can be subtracted to double the modulated component of the interferogram, while in principle cancelling any common mode noise. Alternatively, the two output ports can be configured to simultaneously observe two different wavelength ranges. One disadvantage of the MP is that, in its simplest form, it accepts only one polarization of the incoming beam, which not only reduces the incident flux, but also renders the design sensitive to source polarization. Polarization in the continuum emission could arise from the alignment of dust grains in molecular clouds, or in the line emission from the presence of magnetic fields (e.g. Zeeman effect, Goldreich-Kylafis effect, see \cite{crutcher2012} for a recent review).

\begin{figure}
	\resizebox{\hsize}{!}{\includegraphics{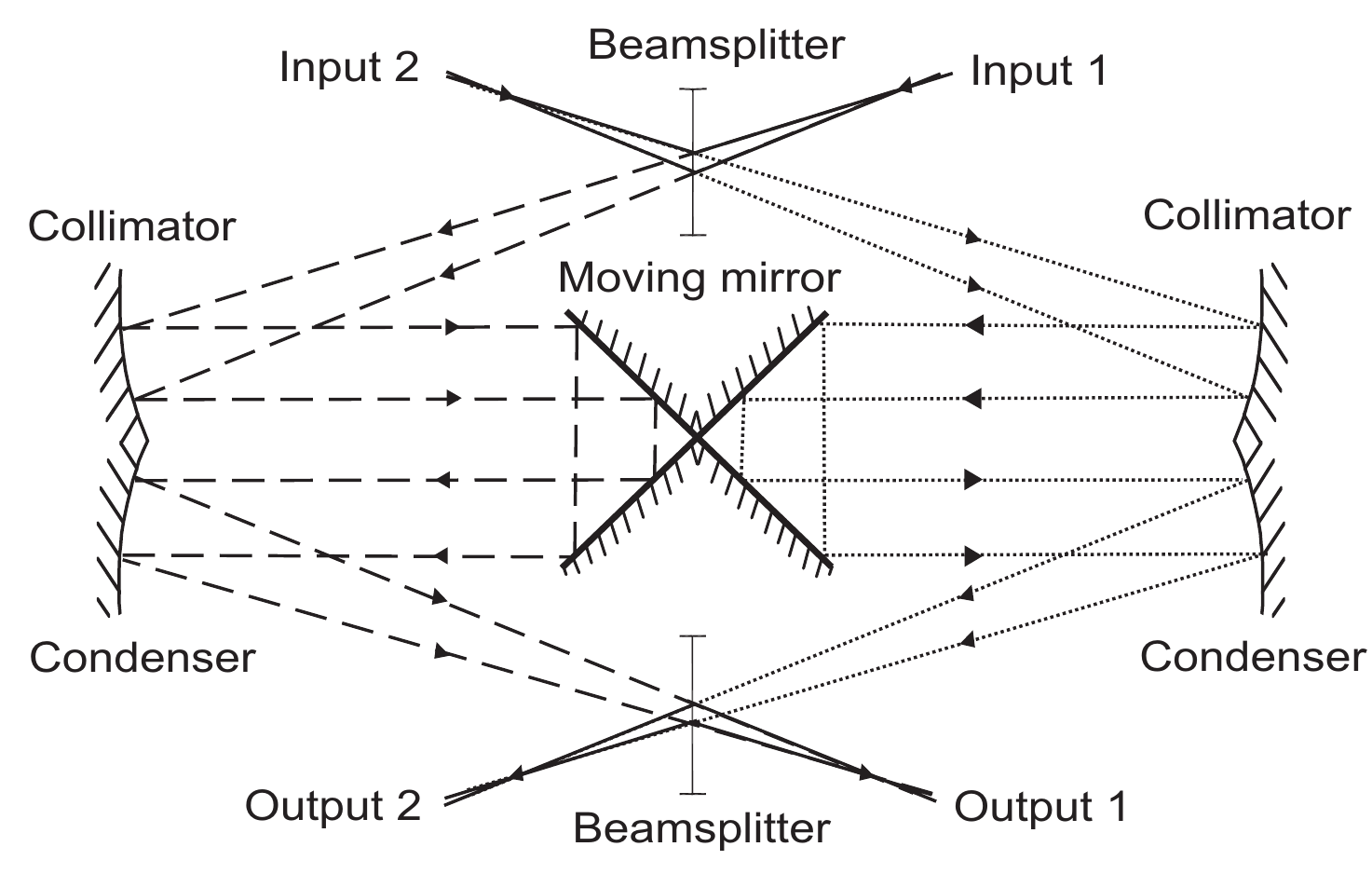}}
	\caption{Optical configuration of the Mach-Zehnder interferometer. After being split by the input beamsplitter, beams from both inputs are collimated and directed to the moving mirror assembly by powered mirrors. A difference in optical path length is introduced in the two beams by a translation of the moving mirror assembly. The collimated beams are focused by the condenser mirrors and recombined by the output beamsplitter; the signals at the two outputs represent complementary interferograms of the difference between the two inputs.}
	\label{fig:MZ}
\end{figure}

The Mach-Zehnder (MZ) interferometer \cite{ade1999,naylor2003a}, shown in Figure~\ref{fig:MZ}, exploits recent advances in the development of intensity beamsplitters \cite{ade2006}, which provide both high and uniform efficiency over a broad spectral range. Like the MP design, the MZ provides access to the two input and two output ports, which can be configured in a variety of ways as discussed above. The principal advantage of the MZ is that all of the incident radiation from the astronomical source reaches the detector arrays. However this comes at the price of increased optical path, and thus component size within the MZ. In order to reduce the size of the beamsplitters, which are typically the most demanding component of any interferometer, the collimating and focusing optics are placed within the interferometer as shown in Figure~\ref{fig:MZ}.

While the separation of the two input and two output ports is a necessary condition to eliminate unwanted contributions to the measured signal from the detector environment, it is not sufficient. Each beamsplitter has a small but non-zero emissivity. It is thus possible for the first beamsplitter to emit light from within its volume, which travels through both arms of the interferometer, that results in an interference signal that is readily detectable with the sensitivity of modern detectors \cite{spencer2011}. In essence, this is the classic double-slit experiment of quantum physics \cite{brukner2002}. Since interference arising from beamsplitter emission is fundamentally unavoidable, it is essential to cool the instrument to cryogenic temperatures. This also reduces contributions to the measured signal from stray light within the instrument, either modulated or unmodulated, which is particularly difficult to control at long wavelengths due to the effects of diffraction.

\section{Examples of Mach-Zehnder iFTS}
For over 30 years the Astronomical Instrumentation Group (AIG) at the University of Lethbridge has developed Fourier transform spectrometers for astronomical research at far-infrared wavelengths. These systems evolved from the classical Michelson interferometer \cite{naylor1986} to the Martin Puplett \cite{naylor1994} culminating in the Mach-Zehnder \cite{ade1999} design. The following sections compare and contrast three MZ interferometers in which the AIG is heavily involved: The spectrometer of the SPIRE instrument on board the {\it Herschel} Space observatory launched in May 2009 \cite{griffin2010}; FTS-2, the spectrometer developed for use with the SCUBA-2 camera currently being commissioned at the JCMT \cite{naylor2006}; and the spectrometer being developed for the SAFARI instrument \cite{roelfsema2012} on the JAXA led SPICA mission. 

\subsection{{\it Herschel}-SPIRE}

\begin{figure}
	\resizebox{\hsize}{!}{\includegraphics{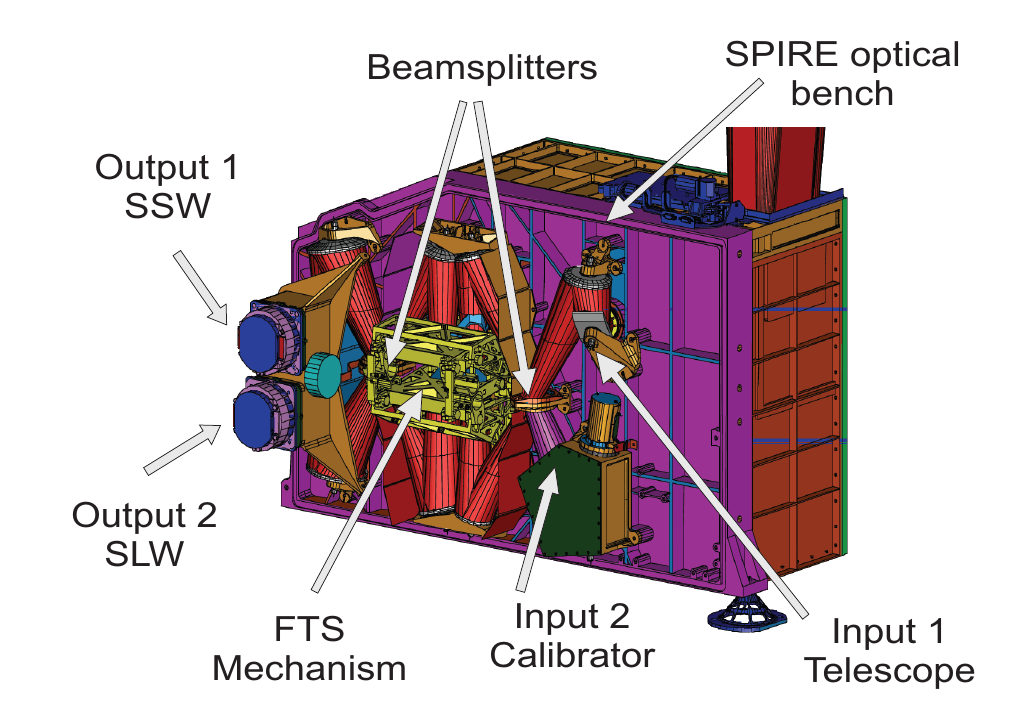}}
	\caption{A schematic of the {\it Herschel}-SPIRE MZ Fourier transform spectrometer. Light from the {\it Herschel} telescope enters through input 1. The temperature of the optical bench is $\sim$4~K; the short and long wavelength detector arrays (SSW, SLW) are at $\sim$0.3~K. Figure adapted from Griffin et al. \cite{griffin2008}.}
	\label{fig:SPIRE}
\end{figure}

The Spectral and Photometric Imaging Receiver (SPIRE) is one of three scientific instruments onboard the European Space Agency's {\it Herschel} Space Observatory launched on 14 May 2009 \cite{pilbratt2010}. Canadian participation in SPIRE is led by the AIG. SPIRE contains a three-band imaging photometer operating at 250, 350 and 500~\micron, and an imaging Fourier Transform Spectrometer (FTS), which covers the wavelength range of 194--671~\micron\ (447--1545 GHz; 14.9--51.5~\pcm) \cite{griffin2010}. The SPIRE instrument, shown in Figure~\ref{fig:SPIRE}, is approximately 700 $\times$ 400 $\times$ 400~mm in size and is supported on the $\sim$4~K {\it Herschel} optical bench by thermally insulating mounts. The photometer and spectrometer are mounted on opposite sides of the SPIRE optical bench, which contains the optics, detector arrays (three for the photometer, and two for the spectrometer), an internal $^3$He cooler to provide the required detector operating temperature of $\sim$0.3~K, filters, mechanisms, internal calibrators, and housekeeping thermometers. Both the photometer and spectrometer have cold pupil stops conjugate with the {\it Herschel} secondary mirror, which is the telescope system pupil, and defines an effective 3.29~m diameter primary aperture. All five detector arrays use hexagonally close-packed, conical feedhorn-coupled spider web neutron transmutation doped Germanium bolometers \cite{turner2001}. The bolometers are AC biased with frequency adjustable between 50 and 200 Hz, avoiding 1/$f$ noise from the cold JFET readout.

The spectrometer is of the Mach-Zehnder configuration and uses two broadband intensity beam splitters \cite{ade2006} to provide spatial separation of the input and output ports. One input port views a 2.6\arcmin\ diameter field of view on the sky and the other an on-board reference source. Two bolometer arrays at the output ports cover overlapping bands of 194--313~\micron\ (SSW) and 303--671~\micron\ (SLW). As with any FTS, each scan of the moving mirror produces an interferogram in which the spectrum of the entire band is encoded with the spectral resolution corresponding to the maximum mirror travel. Specifications of the SPIRE spectrometer are summarized in Table~\ref{t:SPIRE}. A detailed description of the SPIRE spectrometer is given elsewhere \cite{griffin2010}.

\begin{table*}[!htb]
\caption{Summary of the SPIRE instrument specifications.}
\label{t:SPIRE}
\begin{tabular}{r l}
\hline\hline
Spectral range	& 14.9--51.5~\pcm; 447--1545~GHz; 671--194~\micron \\
Maximum spectral resolution	& 0.04~\pcm; $\sigma/\Delta\sigma = 384$ at 650~\micron, 1250 at 200~\micron \\
Spectral resolution modes	& Low (0.83~\pcm), high (0.04~\pcm) \\
\hline
Effective telescope diameter	& 3.29 m \\
Beam diameter at rooftop mirror & $\sim$20 mm \\
Detector optics				& Feedhorn coupled \\
Beam shape				& Gaussian and Multi-moded \\
Angular resolution			& 17--42\arcsec\ (FWHM equivalent beam) \\
Detector arrays				& 37 (SSW) + 19 (SLW) NTD Ge Bolometers \\
Instantaneous field of view	& Circular, 2.6\arcmin, partially vignetted \\
Imaging					& 2$F\lambda$ spacing; Nyquist sampling with 4$\times$4 jiggle pattern \\
\hline
Size						& 0.7 $\times$ 0.4 $\times$ 0.4 m \\
Mass					& 91 kg (including three-band photometer) \\
\hline
Dominant noise source		& $\sim$80~K primary telescope dish \\
Operational temperature		& 4.5~K (optical bench), 0.3~K (detectors), $\sim$80~K (primary mirror) \\
NEFD\tablefoot{a} per pixel per 0.04~\pcm 	& $\sim$1000 mJy (5-$\sigma$, 1 hour)\tablefoot{b} \\
Spectral line sensitivity		& $\sim$$10^{-17}$~W\,m$^{-2}$ (5-$\sigma$, 1 hour) \\
\hline\hline
\end{tabular} \\
\small
\tablefoot{a} Noise equivalent flux density.\\
\tablefoot{b} A milliJansky (mJy) is a unit of flux density commonly used in astronomy; $1~\mathrm{Jy} = 10^{-26}\ \mathrm{W\,m^{-2}\,Hz^{-1}}$.
\normalsize
\end{table*}

\subsection{SCUBA-2 FTS-2}
SCUBA-2 \cite{holland2013} is the first common-user instrument for submillimetre astronomy employing a large format superconducting Transition Edge Sensor (TES) bolometer array, and is currently operating on the James Clerk Maxwell Telescope (JCMT) on Mauna Kea, Hawaii. In order to achieve sky background limited sensitivity under all observing conditions in the narrow 450 and 850~\micron\ atmospheric transmission windows, the SCUBA-2 bolometers had to be designed for low phonon noise-equivalent power and large saturation power, which led to the choice of MoCu bilayer TES devices operating at $\sim$100~mK \cite{holland2013}. The technological innovation of large `CCD-like' detector arrays with atmospheric background noise limited performance provides an unprecedented 2 orders of magnitude increase in mapping speed over the previous camera, SCUBA \cite{holland1999}. Advances in mapping speed and sensitivity are necessary for furthering our understanding of galaxy formation and evolution, a field whose study has seen rapid progress as a direct result of the first generation of submillimetre instruments. Two ancillary instruments, a polarimeter (POL-2, \cite{bastien2011}) and imaging spectrometer, have also been developed to further extend the capabilities of SCUBA-2. A Fourier transform spectrometer was selected as the optimal intermediate resolution spectrometer for SCUBA-2. The instrument, named FTS-2 \cite{naylor2003b}, will be primarily a galactic spectrometer (e.g. spectral index mapping of molecular clouds), but will also provide useful information on bright nearby galaxies and planetary atmospheres. FTS-2 thus fills a niche between the dual band SCUBA-2 continuum images and the higher spectral resolution, but smaller pixel count images produced by the JCMT heterodyne facility instrument HARP-B \cite{buckle2009}.

\begin{figure}
	\resizebox{\hsize}{!}{\includegraphics{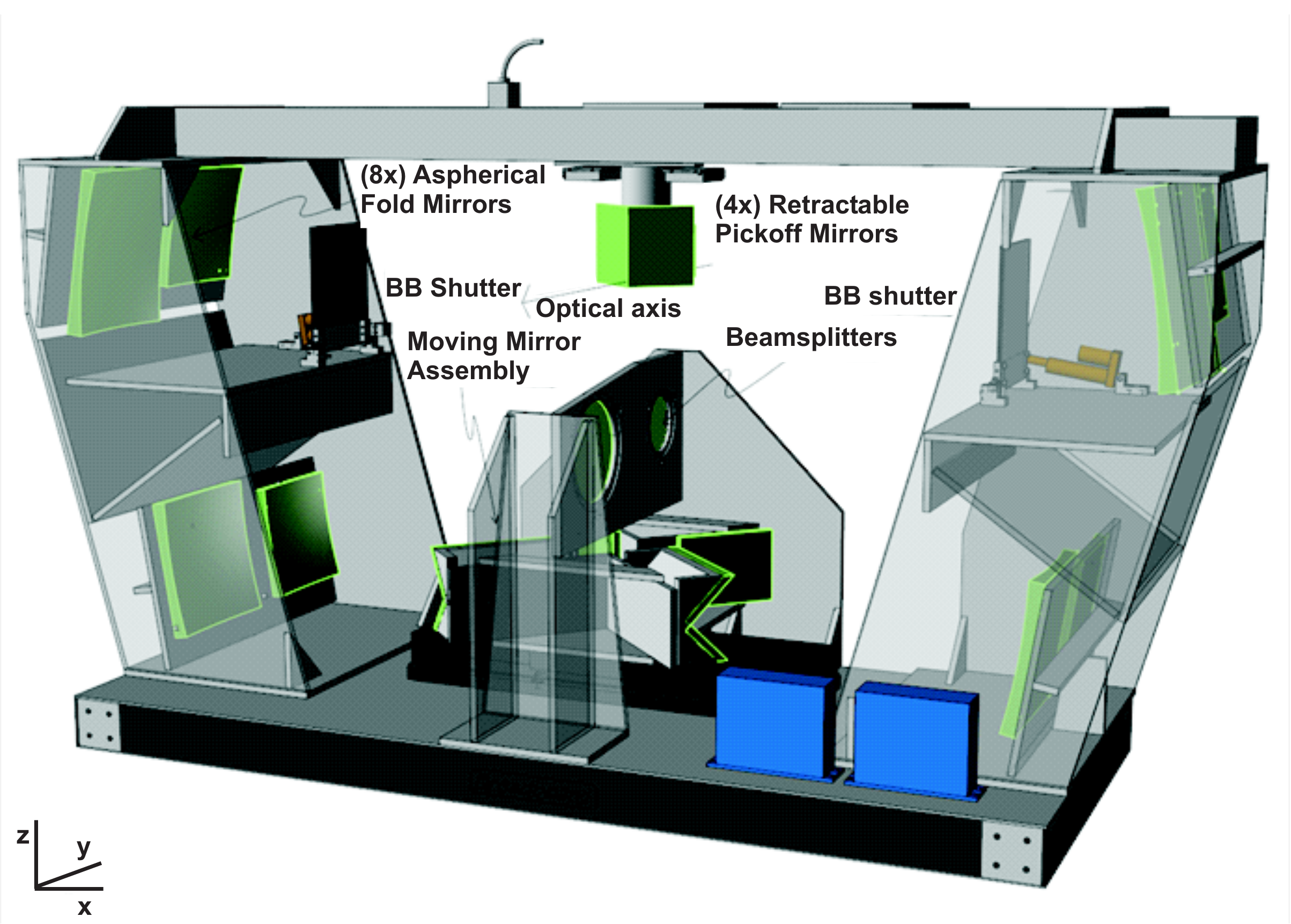}}
	\caption{The FTS-2 instrument, showing the 8 aspherical fold mirrors, the moving back-to-back corner-cube mirror assembly, input and output beamsplitters, two retractable input port shutters, and retractable pickoff mirror assembly. The FTS is mounted perpendicular to the optical axis (which travels along the y axis) at the JCMT Nasmyth focus, midway through the series of SCUBA-2 feed optics.}
	\label{fig:FTS-2}
\end{figure}

\begin{table*}[!htb]
\caption{Summary of the SCUBA-2 FTS-2 instrument specifications.}
\label{t:FTS-2}
\begin{tabular}{r l}
\hline\hline
\multirow{2}{*}{Spectral range}	& 850~\micron: 11.0--12.3~\pcm; 330--369 GHz; 910--813~\micron \\
						& 450~\micron: 21.1--23.5~\pcm; 633--705 GHz; 474--426~\micron \\
Maximum spectral resolution	& 0.006~\pcm; $\sigma/\Delta\sigma = \sim$$1950$ at 850~\micron,  $\sim$$3700$ at 450~\micron \\
Spectral resolution modes	& Low (0.1~\pcm), high (0.006~\pcm) \\
\hline
Effective telescope diameter	& 15 m \\
Beam diameter at rooftop mirror & $\sim$140 mm \\
Detector optics 			& Direct imaging, beam defined by cold-stop \\
Beam shape				& Gaussian \\
Angular resolution			& $\sim$7\arcsec\ FWHM at 450~\micron, $\sim$14\arcsec\ at 850~\micron \\
Detector arrays				& $\sim$800 TES bolometers $\times$ 2 output ports $\times$ 2 wavebands = 3200 \\
Instantaneous field of view	& Circular, $\sim$3\arcmin\ at low resolution \\
Imaging					& 0.5$F\lambda$ spacing at 850~\micron, $F\lambda$ spacing at 450~\micron \\
\hline
Size						& $\sim$2 $\times$ 0.6 $\times$ 1.3 m \\
Mass					& $\sim$600 kg \\
\hline
Dominant noise source		& Atmospheric emission \\
Operational temperature		& 280 K (optical bench), 0.1 K (detectors), $\sim$270 K (primary mirror) \\
NEFD per pixel per 0.006~\pcm & 500 mJy at 850 \micron, 2500 mJy at 450 \micron \\
Spectral line sensitivity		& $\sim10^{-18}$~W\,m$^{-2}$ (5-$\sigma$, 1 hour) \\
\hline\hline
\end{tabular}
\end{table*}

A schematic of the FTS-2 instrument is shown in Figure~\ref{fig:FTS-2}. The Mach-Zehnder configuration [14] is evident, which is made possible by the novel intensity beam dividers. FTS-2 is configured so that both interferometer input ports are placed on the sky and both output ports imaged by the SCUBA-2 arrays. The instantaneous differencing of the two input ports cancels any common-mode atmospheric emission variation, which occurs on faster timescales than the interferogram acquisition time and is the dominant source of noise. Recording the complementary interferograms in both output ports allows for gaps in the image due to dead pixels to be filled in. 

The telescope beam is intercepted at the intermediate \mbox{Nasmyth} image plane by a retractable pickoff mirror assembly, which diverts the light through the FTS and then passes the original image back to the SCUBA-2 feed optics. Since the FTS was designed after the SCUBA-2 optics were finalized, compromises in the field of view and resolution had to be made in order to fit within the available space. Large aspherical diamond-turned aluminum mirrors are required in the interferometer arms to provide a pupil image at the moving mirrors, and to reimage the curved input image plane at the output ports. Corner-cube retroreflectors are necessary in the interferometer moving mirror assembly to compensate for the vertical folding of the interferometer arms. FTS-2 is currently being commissioned at the JCMT. The key specifications are summarized in Table~\ref{t:FTS-2}. A detailed description of the spectrometer is given elsewhere \cite{gom2010}.

\subsection{SPICA-SAFARI}

\begin{figure*}
	\resizebox{\hsize}{!}{\includegraphics{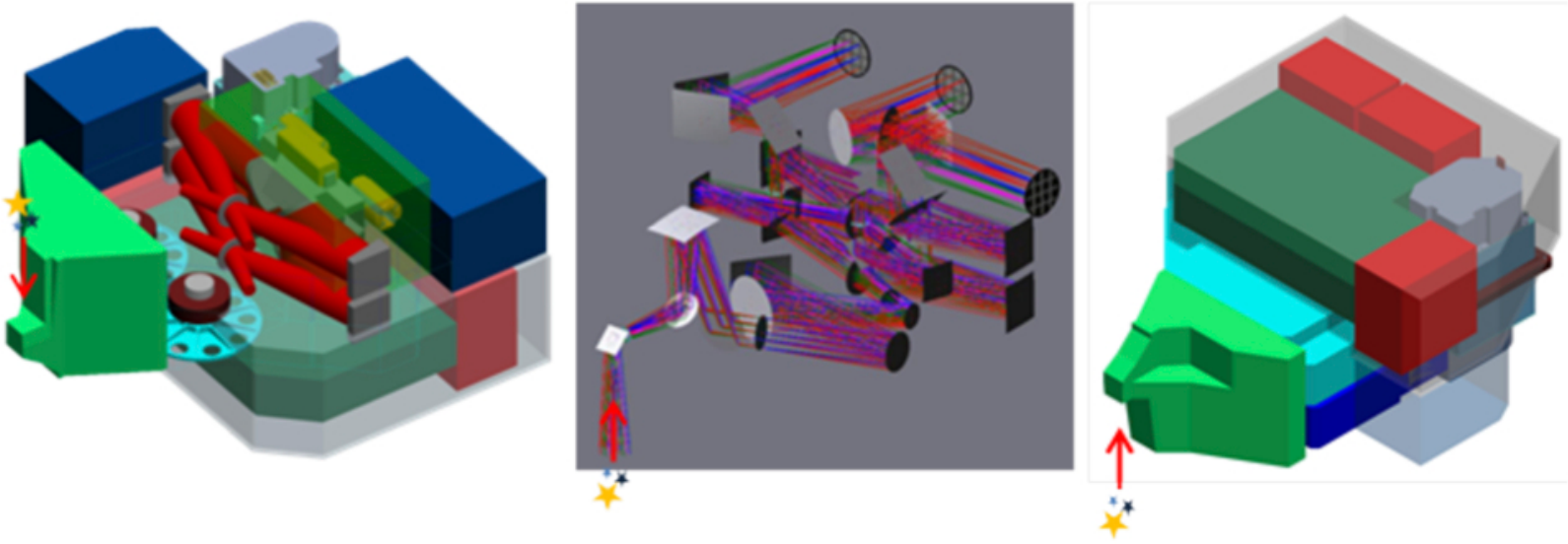}}
	\caption{Optical design of the SPICA-SAFARI FTS. The left panel shows the Focal Plane Unit (FPU) viewed from the top; the infrared radiation from the telescope enters at the far left from the top as indicated by the arrow. Also shown are the two filter wheels, the optical beam path through the FTS (in red), and the FTS mechanism. The right panel shows the FPU from the bottom side. The light green box contains the input optics, the red boxes represent the Focal Plane Assemblies, the blue boxes house the filters for the readout wiring, and the dark green box houses the camera bay optics where the FTS output signal is directed to the three detector arrays. The middle panel (same orientation as the right panel) shows the optical path through the interferometer from the pick-off mirror on the bottom left to the three detector arrays on the top right. Figure courtesy of Roelfsema et al. \cite{roelfsema2012}.}
	\label{fig:SAFARI}
\end{figure*}

\begin{table*}[!htb]
\caption{Summary of the SPICA-SAFARI FTS instrument specifications.}
\label{t:SAFARI}
\begin{tabular}{r l}
\hline\hline
Spectral range				& 47--294~\pcm; 1428--8817 GHz; 210--34 \micron \\
Maximum spectral resolution	& 0.05~\pcm; $\sigma/\Delta\sigma = 2000$ at 100~\micron \\
Spectral resolution modes	& Photometry ($\sigma/\Delta\sigma$ = 3), low ($\sigma/\Delta\sigma$ = 150--200), high ($\sigma/\Delta\sigma$ = 2000) \\
\hline
Effective telescope diameter	& $\sim$3 m \\
Beam diameter at rooftop mirror &  33 mm \\
Detector optics				& Feedhorn coupled \\
Beam shape				& Determined by feedhorn \\
Angular resolution			& Diffraction limited above 40 \micron\ ($\sim$3.5\arcsec\ at 40 \micron, $\sim$18\arcsec\ at 210 \micron) \\
Detector arrays				& 43$\times$43 (short wave) + 34$\times$34 (medium wave) + 18$\times$18 (long wave) = 3329 TES detectors \\
Instantaneous field of view	& Circular, $\sim$2\arcmin \\
Imaging					& 0.5$F\lambda$ spacing at centre of each band \\
\hline
Size						& TBD \\
Mass					& $<$50 kg \\
\hline
Dominant noise source		& Goal zodiacal background, photon noise limited \\
Operational temperature		& 4.5 K (optical bench), 50 mK (detectors), 6 K (primary mirror) \\
NEFD per pixel per 0.05~\pcm & 15 mJy (5-$\sigma$, 1 hour) \\
Spectral line sensitivity		& few $10^{-19}$~W\,m$^{-2}$ (5-$\sigma$, 1 hour) \\
\hline\hline
\end{tabular}
\end{table*}

As noted above, despite its low emissivity, thermal emission from the passively cooled {\it Herschel} telescope produces a background flux in the 20--500 \micron\ range many orders of magnitude greater than that from galactic cirrus clouds, the zodiacal light and the 3~K cosmic microwave background. The next advance in far-infrared space astronomy requires a cold telescope to explore the cold universe. 

It is this unique low thermal background environment that the Japanese led Space Infrared telescope for Cosmology and Astrophysics (SPICA) mission will provide in the early 2020's. With its 3~m class primary mirror cooled to $\sim$6~K, the thermal emission from the SPICA telescope will be 6 orders of magnitude less than on JWST \cite{gardner2006} or {\it Herschel} \cite{pilbratt2010}. A new generation of superconducting transition edge sensor detectors will achieve sensitivities 2 orders of magnitude better than existing space borne detectors to explore this low background environment. A consortium of European and Canadian scientists, which includes the authors, are developing a sensitive far infrared imaging spectrometer, SAFARI -- the SpicA FAR infrared Instrument. SAFARI will obtain fully spatially sampled spectra over the range 34--200 \micron\ and a 2\arcmin$\times$2\arcmin\ field enabling it to detect and measure, in a 900 hour integration, spectra of approximately one thousand times more extragalactic sources than the PACS instrument \cite{poglitsch2010}, currently the most sensitive space borne spectrometer in this wavelength range.

Building on the success of {\it Herschel}, the SAFARI consortium has selected an imaging Fourier transform spectrometer of the Mach-Zehnder design to achieve the spectral imaging goals. A schematic of the SAFARI instrument is shown in Figure~\ref{fig:SAFARI}. The design is described elsewhere \cite{roelfsema2012}. To cover the \mbox{34--210~\micron} range with adequate spatial and spectral resolution, the interferometer employs three detector arrays. One output port is directed to the long wavelength detector channel (110--210 \micron), a dichroic placed after the second output port splits the signal into the medium (60--110 \micron) and short \mbox{(34--60 \micron)} waveband. Full resolution spectra (corresponding to $\sigma/\Delta\sigma$ of 2000 at 100 \micron) require a 35~mm mechanical displacement of the moving mirror assembly, accomplished in a time of $\sim$2 minutes. 

\begin{figure}
	\resizebox{\hsize}{!}{\includegraphics{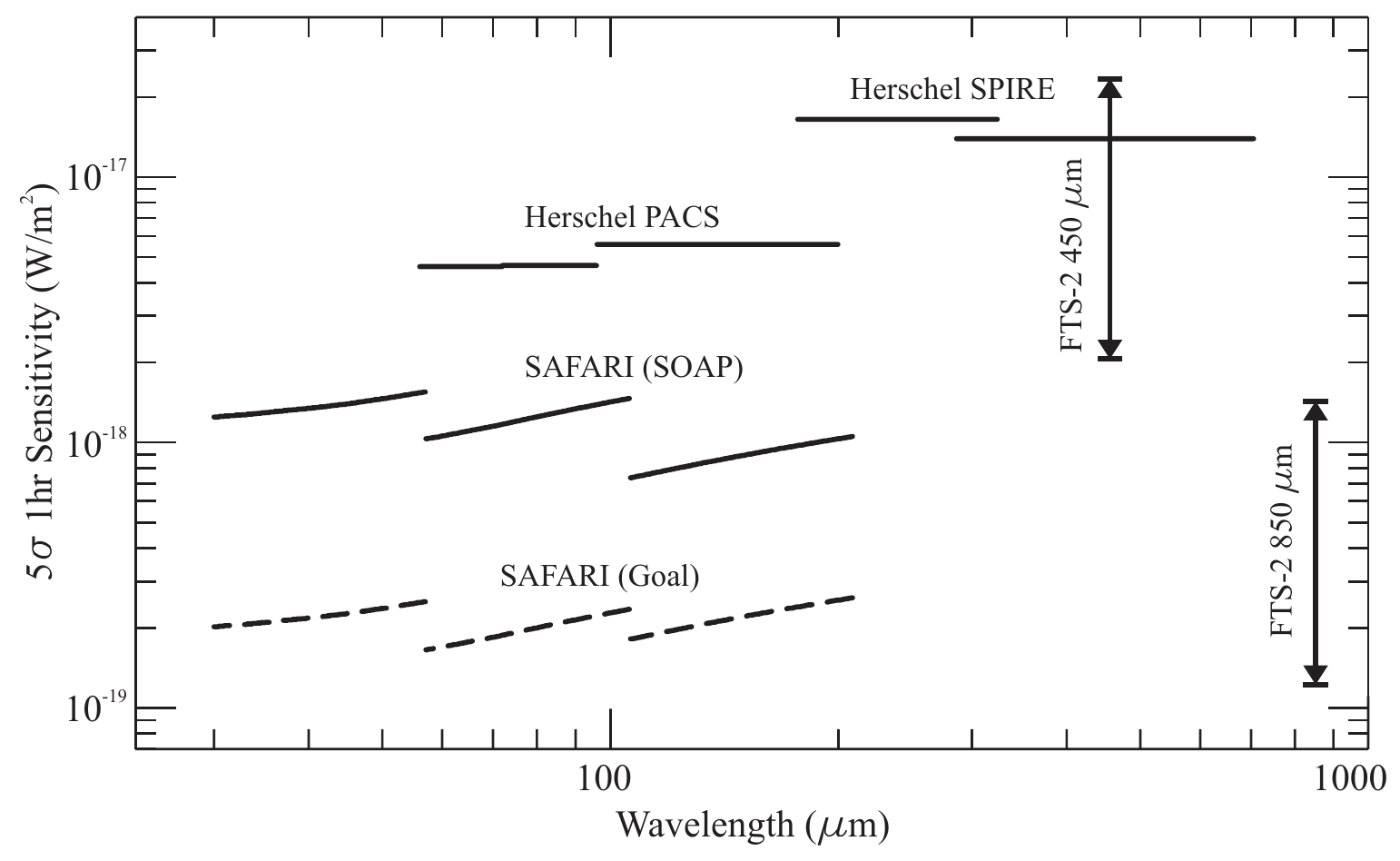}}
	\caption{Comparison of the measured and predicted performances, defined as the 5-$\sigma$ spectral line sensitivity attainable in one hour of observing time, of the {\it Herschel}-SPIRE, SCUBA-2 FTS-2 and SPICA-SAFARI Mach-Zehnder interferometers. For SAFARI, the state-of-the-art detector performance (SOAP) assumes photoconductor detectors, while the goal performance assumes TES bolometers. \cite{naylor2011} }
	\label{fig:comparesens}
\end{figure}

The key specifications of the SAFARI instrument are summarized in Table~\ref{t:SAFARI}. The measured and predicted sensitivities of the {\it Herschel}-SPIRE, SCUBA-2 FTS-2 and SPICA-SAFARI imaging Fourier transform spectrometers are compared in Figure~\ref{fig:comparesens}.

\section{FTS Spectral Analysis}

\begin{figure*}
	\resizebox{\hsize}{!}{\includegraphics{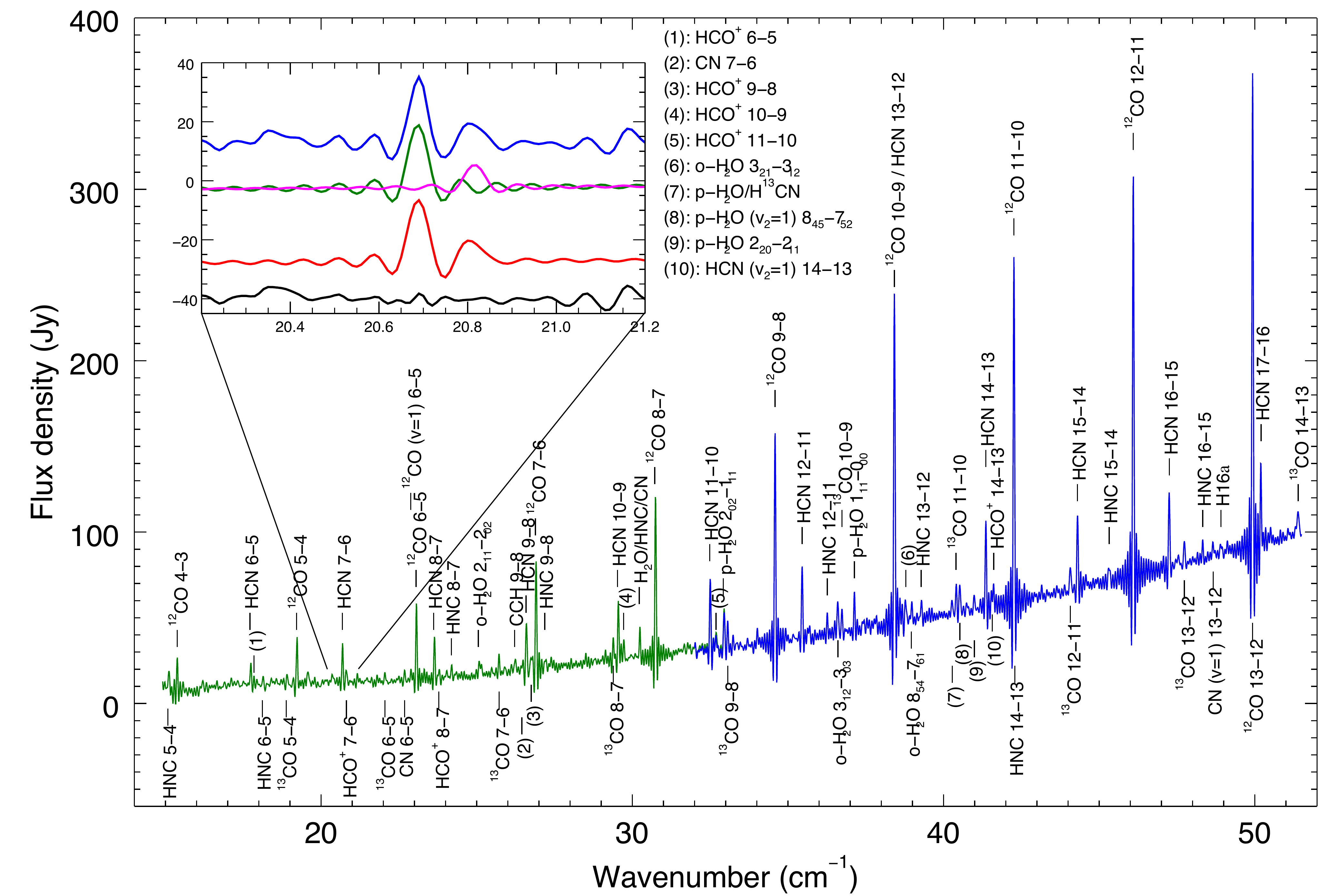}}
	\caption{{\it Herschel}-SPIRE spectrum of AFGL618 with identified transitions labeled. The inset diagram illustrates the fitting of two blended spectral lines in the observed spectrum (blue). Two individual sinc profiles are shown in green and magenta, their superposition in red, and the residual in black, all displaced vertically for clarity.}
	\label{fig:AFGL618spectrum}
\end{figure*}

In this section we show how a {\it Herschel}-SPIRE spectrum is analyzed to derive astrophysical parameters of the source under study. The analysis is generic and can be applied to any spectrum obtained with a Fourier transform spectrometer. Figure~\ref{fig:AFGL618spectrum} shows the spectrum of one of the SPIRE calibration targets, the post-AGB (asymptotic giant branch) star AFGL618. This object is compact relative to the {\it Herschel} telescope beam and is known for the abundance of chemical species present in its molecular wind \cite{herpin2000,wesson2010}; it is thus an excellent point source for calibrating the broadband spectrometer. Spacecraft pointing errors are potentially troublesome for a Fourier transform spectrometer as they can produce spectral artefacts which result in an additional source of error in the retrieved spectral parameters. Repeated observations of calibration targets, such as AFGL618, throughout the mission, however, yield integrated line intensities that vary by less than 4\%, which confirms that the low {\it Herschel} pointing uncertainty ($\sim$2\arcsec) is not a dominant source of error. The observation shown in Figure~\ref{fig:AFGL618spectrum} was obtained with a 10-minute on-source exposure [{\it Herschel} observation ID 1342242592, 2012 March 10th]. The spectrum is the result of standard processing with the SPIRE FTS pipeline (v9) provided by the {\it Herschel} Science Centre and the SPIRE instrument control centre \cite{ott2010}. 

The spectrum reveals a baseline of continuum emission, which is due to dust grains. Superimposed on this continuum is an array of sinc-shaped spectral lines that arise from gaseous components in the emitting region. We have developed a line fitting tool for FTS spectra \cite{naylor_inprep}. The algorithm performs an iterative fit to the entire spectrum with a superposition of a polynomial (for the continuum) and a number of spectrally unresolved lines. Since the intrinsic width of the spectral lines in AFGL618 is much smaller than the spectral resolution of the SPIRE FTS, the line shape is governed by the instrumental sinc line shape, which has a width of 0.04~\pcm, determined from the maximum optical path difference in the interferogram. The area (integrated line intensity) and peak position (central frequency) for each line are returned by the line fitter. By comparing the line frequencies with tabulated values of known interstellar gas species (JPL \cite{pickett1998}, CDMS \cite{mueller2005}), lines can be identified, and are indicated by the labels in Figure~\ref{fig:AFGL618spectrum}. It can be seen that the rich spectrum of AFGL618 contains lines from molecular, atomic and ionic species such as CO, H$_2$O, H, HCN, HCO$^+$ and CCH. 

\begin{figure}
	\resizebox{\hsize}{!}{\includegraphics{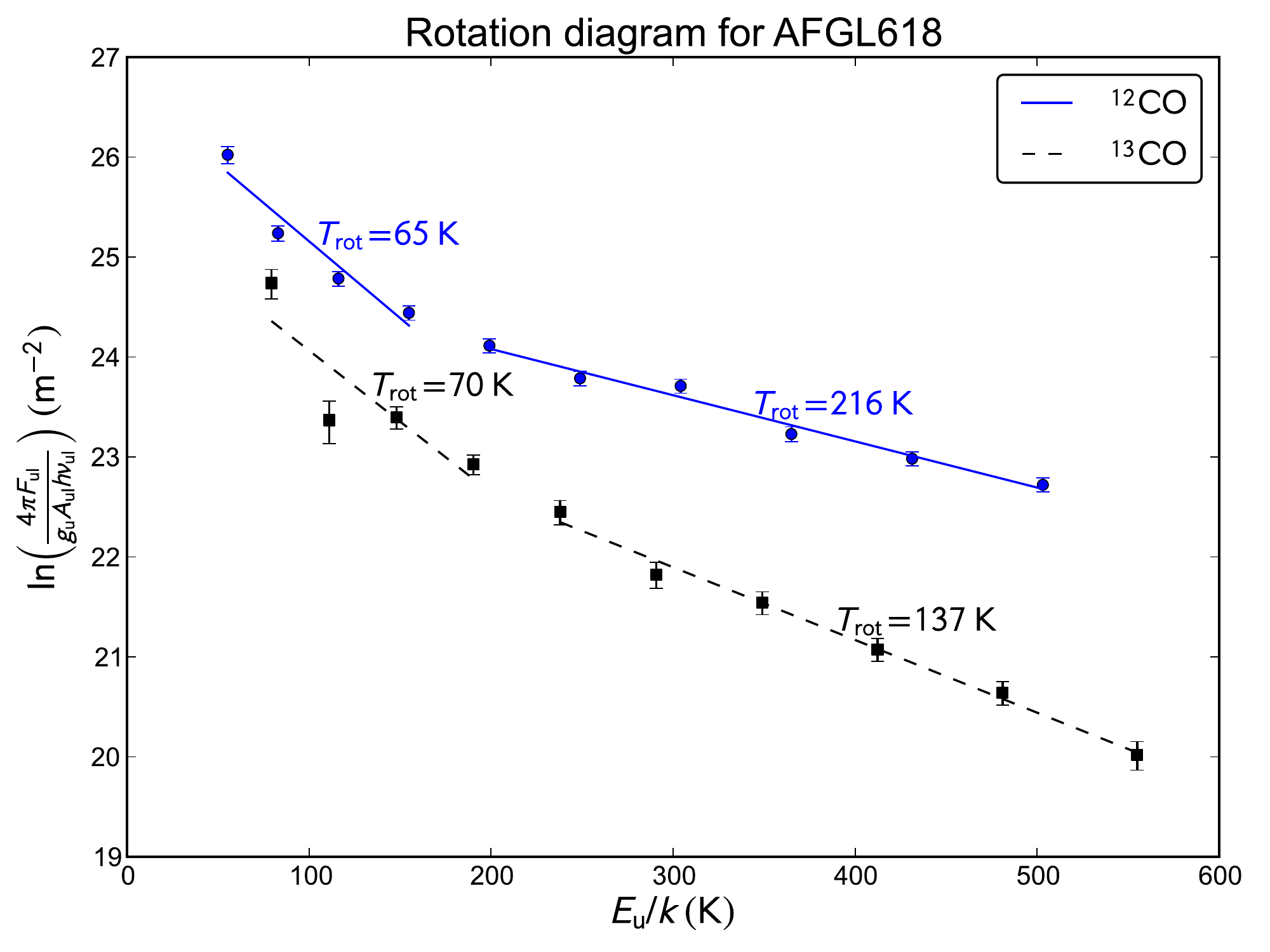}}
	\caption{Rotation diagram for $^{12}$CO (blue circle points and solid line) and $^{13}$CO (black square points and dashed lines) in AFGL618. Points represent scaled measured line intensities ($y$ in Equation~[\ref{eq:rotatdiag}]) of, from left to right, the $J$=4--3 up to 13--12 transitions ($^{12}$CO) and 5--4 up to 14--13 ($^{13}$CO), with formal fitting uncertainties shown by vertical error bars. Straight lines indicate the linear regression fits to the data points. }
	\label{fig:rotatdiag}
\end{figure}

Molecular species such as CO have a ladder of rotational transitions in the spectral range covered by SPIRE as seen in Figure~\ref{fig:AFGL618spectrum}. Observing several transitions with varying upper level energies, $E_\mathrm{u}$, allows one to derive physical parameters for the emitting gas. A so-called `rotation diagram' \cite{turner1991,goldsmith1999} is constructed under the assumptions that (a) the gas can be represented by a single temperature ($T_\mathrm{X}$ or $T_\mathrm{rot}$), (b) the lines are optically thin, (c) local thermodynamic equilibrium (LTE) applies, i.e., the excitation balance of the molecule is dominated by collisional excitations, and (d) the emitting region fills the telescope beam uniformly. In this case, the integrated line intensity for each transition from upper level `u' to lower level `l' is directly proportional to $N_\mathrm{u}$, the number of molecules in state `u', which is itself a function of $T_\mathrm{X}$:
\begin{eqnarray}
F_\mathrm{ul} = & \frac{\mathrm{(rate\ of\ transitions)}\times\mathrm{(energy/photon)}}{\mathrm{area}} \times N_\mathrm{u} \nonumber \\ 
  = & \frac{A_\mathrm{ul}h\nu_\mathrm{ul}}{4\pi d^2} \frac{N_\mathrm{tot}}{Q} g_\mathrm{u} \mathrm{e}^{-E_\mathrm{u}/k T_\mathrm{X}}
\end{eqnarray}
with $h$ and $k$ the Planck and Boltzmann constants, respectively, $A_\mathrm{ul}$ the Einstein coefficient for spontaneous emission, $\nu_\mathrm{ul}$ the frequency of the emitted photon, $d$ the distance to the object, $N_\mathrm{tot}$ the total number of (CO) molecules summed over all states, $Q$ the partition function, and $g_\mathrm{u}$ the statistical weight of level u.  Rearranging this equation yields:
\begin{equation}
\underbrace{\ln \left( \frac{4\pi F_\mathrm{ul}}{g_\mathrm{u} A_\mathrm{ul} h \nu_\mathrm{ul}} \right)}_{y} = \underbrace{-\frac{1}{T_\mathrm{X}}}_{m} \underbrace{\frac{E_\mathrm{u}}{k}}_{x} + \underbrace{\ln \left( \frac{N_\mathrm{tot}}{Q d^2} \right)}_{c} .
\label{eq:rotatdiag}
\end{equation}
A graph of $y$ versus $x$ is known as a `rotation diagram'. The slope $m$ of a linear fit to the data points (straight lines in Figure~\ref{fig:rotatdiag}) yields the gas temperature, while the intercept on the $y$ axis ($c$) yields the total number of molecules along the line of sight. 

It can be seen from Figure~\ref{fig:rotatdiag} that the emitting region in AFGL618 cannot be described by a single temperature. Further, since the ratio of the $^{12}$CO/$^{13}$CO line fluxes varies, the $^{12}$CO lines are not optically thin and a full analysis requires detailed modelling of the structure of the source. Since optical depth effects have a greater influence on the vertical offset of the rotation diagram than on its slope, it is still possible to derive meaningful rotational temperatures by invoking two temperature components. The temperature values indicated in Figure~\ref{fig:rotatdiag} are similar to those derived by Wesson et al. \cite{wesson2010}, whose work is based on a SPIRE spectrum of the same source, but having a longer exposure. Even in this example of a calibration source it can be seen that at least two temperature components are required to fit the observed data.

The rotation diagram approach outlined above can be adapted to ÔcorrectÕ for effects of optical depth and non-uniform beam filling \cite{goldsmith1999}. In many interstellar environments, the assumption that the medium is isothermal and in LTE fails and one must resort to a more detailed modeling of the emitting region. In general, such methods rely on iteration to determine interdependent parameters: spatial distributions of dust temperature, gas temperature, molecular excitation balance and in some cases chemical balance \cite{vandertak2011}.

\begin{figure}
	\resizebox{\hsize}{!}{\includegraphics{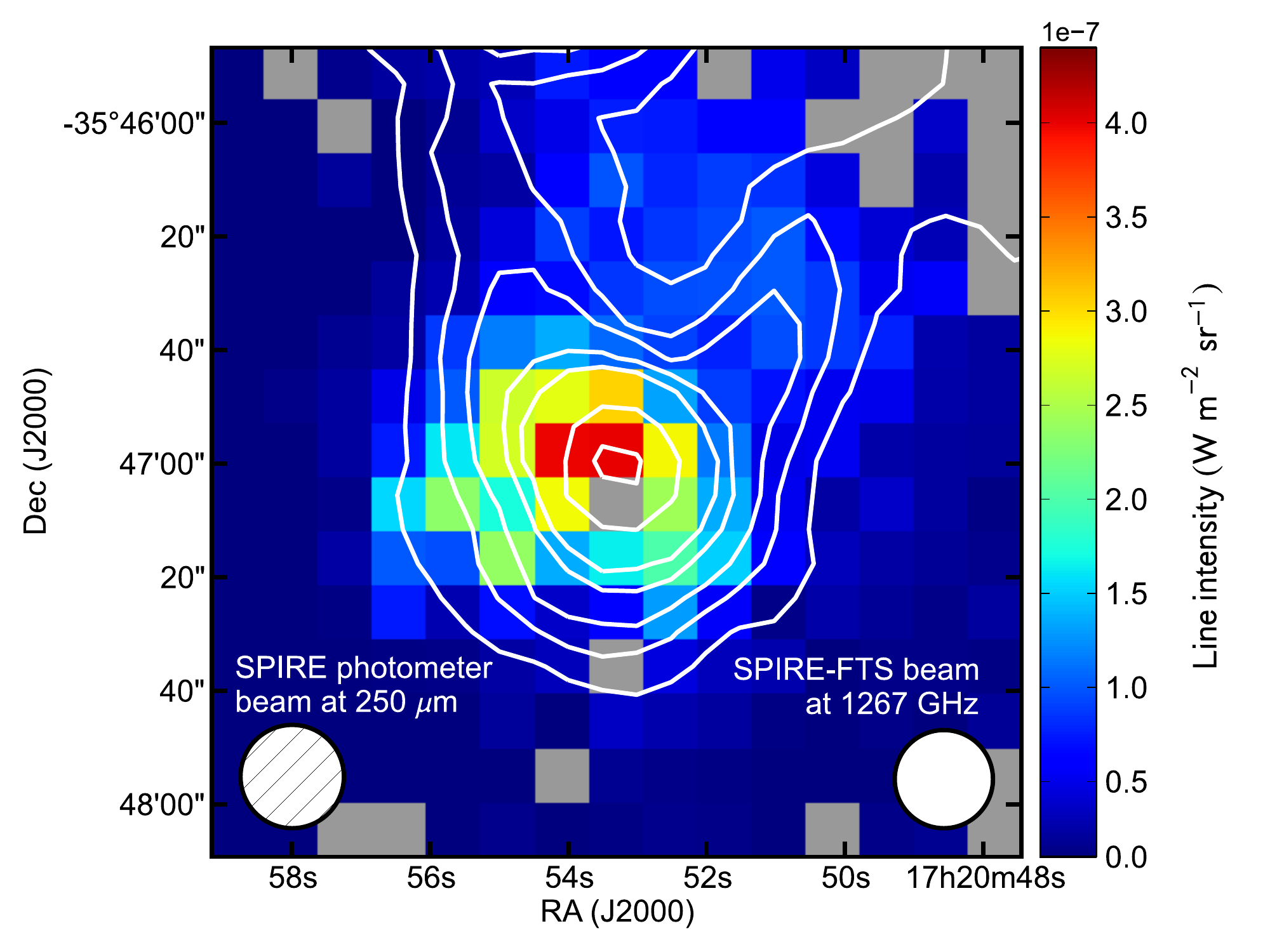}}
	\caption{Integrated line intensity of CO $J$=11--10 toward the high-mass star-forming region NGC6334I (colour scale), observed with {\it Herschel}-SPIRE FTS. White contours indicate the distribution of dust continuum emission measured in the 250~\micron\ band of SPIRE's broadband photometer.  }
	\label{fig:SPIREmap}
\end{figure}

The power of the Mach-Zehnder instruments described in this paper lies in their ability to provide astronomical imaging spectroscopy. A spectrum similar to that shown in Figure~\ref{fig:AFGL618spectrum} is recorded for each pixel in the detector array. To illustrate the spectroscopic imaging capability of the SPIRE FTS, we present observations of the massive star-forming region known as NGC6334I \cite{hunter2006,qiu2011}. The {\it Herschel} observation, part of the {\it Herschel} Key Program ``Evolution of Interstellar Dust" \cite{abergel2010} [{\it Herschel} observation IDs 1342214827 (2011 Feb 26$^\mathrm{th}$) and 1342251326 (2012 Sep 24$^\mathrm{th}$), total 6 hours], produced a spectral cube covering a 2\arcmin$\times$2\arcmin\ region. Analogous to the fitting described above, all the CO lines were fitted in all spatial pixels of this cube. Figure~\ref{fig:SPIREmap} shows the integrated intensity map obtained for one of these CO lines corresponding to the $J$=11--10 transition at 42.26~\pcm\ (1267 GHz). The white contours shown in this figure indicate the distribution of dust continuum emission measured with the SPIRE photometer [{\it Herschel} observation ID 1342239909], as part of the {\it Herschel} Key Program named HOBYS \cite{motte2010}. Combining the two maps allows a direct comparison of the spatial extent of the (cold) dust and, in this case, the hot gas components of emission. Maps from all spectral lines provide astrophysicists with unique tools to compare and contrast the distribution of various chemical species and of different transitions of the same species. Combined with theoretical and numerical modeling, this leads to a better understanding of the physical conditions and chemical history of the regions under study. In addition, the determination of the central position of a spectral line is reliable to within a fraction ($\sim$1/20) of the instrumental line width ($\sigma/\Delta\sigma$ around 1000 for SPIRE FTS), yielding an uncertainty in the line centre equivalent to a velocity error of the order of 10 km\,s$^{-1}$ \cite{naylor2010}.

\section{Summary}

Approximately one half of the radiant energy emitted by the universe falls in the far-infrared/submillimetre spectral range.  Several techniques have been developed to measure and interpret the information conveyed by these low energy photons.  While photometric observations provide information on the morphology of the source under study, the underlying physics is only revealed through spectroscopic measurements. One of the emerging spectroscopic techniques, which is gaining widespread interest, is that of imaging Fourier transform spectroscopy (iFTS). IFTS provides broad spectral coverage at intermediate spectral resolution, for every pixel in the detector bolometer array. Fourier transformation of the measured interferograms produces a spectral data cube having two spatial and one spectral dimension. The designs of the classical Michelson and Martin-Puplett interferometers have been compared and contrasted to the Mach-Zehnder interferometer. Examples of three instruments based on the Mach-Zehnder design have been presented. The techniques for retrieving astrophysical parameters from measured FTS spectra are examined using calibration data obtained with the {\it Herschel}-SPIRE instrument. The paper concludes with an example of imaging spectroscopy obtained with the SPIRE FTS instrument.

\section*{Acknowledgements}
\emph{This paper is dedicated to Professor Arvid A. Schultz, friend and colleague, and founding member of the Physics Department at the University of Lethbridge.} \\
The authors would like to acknowledge support from Alberta Advanced Education and Technology, Canada Foundation for Innovation, Canadian Space Agency (CSA), Natural Science and Engineering Research Council of Canada and the University of Lethbridge. SPIRE has been developed by a consortium of institutes led by Cardiff Univ. (UK) and including Univ. Lethbridge (Canada); NAOC (China); CEA, LAM (France); IFSI, Univ. Padua (Italy); IAC (Spain); Stockholm Observatory (Sweden); Imperial College London, RAL, UCL-MSSL, UKATC, Univ. Sussex (UK); Caltech, JPL, NHSC, Univ. Colorado (USA). This development has been supported by national funding agencies: CSA (Canada); NAOC (China); CEA, CNES, CNRS (France); ASI (Italy); MCINN (Spain); SNSB (Sweden); STFC, UKSA (UK); and NASA (USA). FTS-2 is funded through a CFI international access award. The SAFARI consortium is led by SRON Netherlands Institute for Space Research, Canadian participation is supported by the CSA.

\small
\begin{spacing}{0.9}
\bibliographystyle{plain}

\end{spacing}

\end{document}